\documentclass[]{spie}  

\usepackage{amsmath,amsfonts,amssymb}
\usepackage{graphicx}
\usepackage[colorlinks=true, allcolors=blue]{hyperref}

\graphicspath{{./}{figures/}}

\def\ovi{O~{\sc vi}}
\def\hi{H~{\sc i}}
\def\nv{N~{\sc v}}
\def\civ{C~{\sc iv}}

\title{Mapping Diffuse Emission in Lyman UV band}

\author[a,l]{Li Ji}
\author[a]{Zheng Lou}
\author[b]{Jinlong Zhang}
\author[c]{Keqiang Qiu}
\author[b]{Shuangying Li}
\author[a,l]{Wei Sun}
\author[a,l]{Shuping Yan}
\author[a,l]{Shuinai, Zhang}
\author[a]{Yuan Qian}
\author[d]{Sen Wang}
\author[e]{Klaus Werner}
\author[f]{Taotao Fang}
\author[g]{Tinggui Wang}

\author[e]{J\"urgen Barnstedt}
\author[e]{Sebastian Buntrock}
\author[a,l]{Mingsheng Cai}
\author[h]{Wen Chen}

\author[e]{Lauro Conti}
\author[h]{Lei Deng}
\author[e]{Sebastian Diebold}
\author[c]{Shaojun Fu}
\author[a,l]{Jianhua Guo}
\author[e]{Lars Hanke}
\author[c]{Yilin Hong}
\author[e]{Christoph Kalkuhl}
\author[e]{Norbert Kappelmann}
\author[e]{Thomas Kaufmann} 
\author[a,l]{Shijun Lei}
\author[i]{Fu Li}
\author[j]{Xinfeng Li}
\author[a]{Wei Liu}
\author[k]{Kevin Meyer}
\author[e]{Thomas Rauch}
\author[i]{Ping Ruan}
\author[k]{Daniel M. Schaadt}
\author[e]{Thomas Schanz}
\author[d]{Qian Song}
\author[e]{Beate Stelzer}
\author[b]{Zhanshan Wang}
\author[i]{Jianfeng Yang}
\author[a]{Wei Zhang}

\affil[a]{Purple Mountain Observatory, CAS, Nanjing, China}
\affil[b]{Institute of Precision Optical Engineering, School of Physics Science and Engineering, Tongji University, Shanghai, China}
\affil[c]{National Synchrotron Radiation Laboratory, University of Science and Technology of China, Hefei, China}
\affil[d]{National Astronomical Observatories, CAS, Beijing, China}
\affil[e]{Institute for Astronomy and Astrophysics, Universit\"at T\"ubingen, T\"ubingen, Germany}
\affil[f]{Department of Astronomy, Xiamen University, Xiamen, China}
\affil[g]{Department of Astronomy, University of Science and Technology of China, Hefei, China}
\affil[h]{Innovation of Academy for Microsatellites of CAS, Shanghai, China}
\affil[i]{Xi'an Institute of Optics and Precision Mechanics of CAS, Xi'an, China}
\affil[j]{Technology and Engineering Center for Space Utilization, CAS, Beijing, China}
\affil[k]{Institute of Energy Research and Physical Technologies, Clausthal University of Technology, Clausthal, Germany}
\affil[l]{Key Laboratory of Dark Matter and Space Astronomy, CAS, Nanjing, China}

\authorinfo{Further author information: (Send correspondence to Li Ji)\\Li Ji: E-mail: ji@pmo.ac.cn, Telephone: 86 25 83332162}

\begin{document} 
\maketitle

\begin{abstract}
The CAFE (Census of warm-hot intergalactic medium, Accretion, and Feedback Explorer) and LyRIC (Lyman UV Radiation from Interstellar medium and Circum-galactic medium) have been proposed to the space agencies in China respectively. CAFE was first proposed in 2015 as a joint scientific CAS-ESA small space mission. LyRIC was proposed in 2019 as an independent external payload operating on the Chinese Space Station. Both missions are dedicated to mapping the Lyman UV emission (ionized oxygen (\ovi{}) resonance lines at 103.2 and 103.8 nm, and Lyman series) for the diffuse sources in our Galaxy and the circum-galactic mediums of the nearby galaxies. We present the primary science objectives, mission concepts, the enabling technologies, as well as the current status.
\end{abstract}


\section{INTRODUCTION}
\label{sec:intro}  

The ``Lyman ultraviolet'' (LUV; 91.2--121.6 nm), the least explored window of astrophysical investigations, is rich in the unique physical line diagnostics including neutral hydrogen (\hi{}) Lyman series lines and ionized oxygen (\ovi{}) resonance lines at 103.2 and 103.8 nm. Tumlinson et al. (2012)\cite{Tumlinson2012} and Green et al. (2014)\cite{Green2014} pointed out that with modern photon-counting detectors, the LUV region has been fully covered by the Far Ultraviolet Spectroscopic Explorer (FUSE)\cite{Moos2000}, the Spectroscopy of Plasma Evolution from Astrophysical Radiation (SPEAR)\cite{Edelstein2006a}, and the Hubble Space Telescope (HST)/COS after cycle 20\cite{McCandliss2010}. However, both FUSE and HST/COS are mainly targeting the point sources, not optimized to the larger structures of the diffuse emission from the astronomical sources. SPEAR ever provided the images of LUV diffuse emission line sources in our Galaxy but with relatively low effective area and insufficient effective angular resolution. 

With the characteristic temperature of 10$^4$--10$^6$ K, diffuse gas is the key to understand the circum-galactic medium (CGM) surrounding nearby galaxies, high-velocity clouds (HVCs) in our Galaxy, and superbubbles etc. Aiming at the missions with the exploration size, we have proposed two LUV emission explorers to the space science community in China. One is CAFE (Census of warm-hot intergalactic medium, Accretion, and Feedback Explorer), first proposed in 2015 as a joint scientific CAS-ESA small space mission. The other is LyRIC (Lyman UV Radiation from Interstellar medium and Circum-galactic medium), proposed in 2019 as an independent external payload operating on the Chinese Space Station. We present here the primary science objectives, mission concepts, the enabling technologies, as well as the current status.

\section{Primary Science Objectives}

\subsection{Where are the Missing Baryons?}
In the low redshift universe, only about half of the total baryons in the $\Lambda$CDM cosmological model can be accounted for in stars, cold or warm interstellar matter, hot intracluster gas, and residual photo-ionized intergalactic medium\cite{Nicastro2008,Shull2012}. The remainder is still ``missing'' and under hot debate\cite{Nicastro2018,deGraaff2019,Macquart2020}. Numerical simulations predicted that most of this missing baryonic component should be in the form of warm-hot intergalactic medium (WHIM, $T\sim$10$^5$--10$^7$~K)\cite{Dave2001,Cen2005}, which provide vast filamentary reservoirs of material that continually feed star formation in galaxies. Finding this component and thereby producing a census of the mass budget of the local Universe is crucial to confine the standard cosmological model. These baryons are difficult to observe. The presence of the WHIM has been inferred from the detection of discrete UV absorption features in the spectra of bright background targets\cite{Prochaska2011}. However, such observations yield a relatively small number (few hundred) of random sight-lines, from which physical conditions mostly have to be analyzed statistically; they give no real information on the shapes and flow patterns of the intergalactic gas. With sensitivity down to 500 PU  (Photons~s$^{-1}$~cm$^{-2}$~sr$^{-1}$), CAFE will detect and map the \ovi{} doublet emission 
from the denser and warmer parts of WHIM (i.e. in the vicinity of galaxies), as well as carry out a census of baryons in about 200 galaxy halos, which would critically test the physics in the hydrodynamic simulations.

\subsection{How do accretion and feedback affect galaxy evolution?}

Accretion and feedback are two of the most important and most poorly understood topics of present-day studies of galaxy evolution. Direct observation of gas accretion onto galaxies remains sparse due to the faintness of the measurable signals and the limited sensitivities of current observatories \cite{Putman2012}. \hi{} Ly$\alpha$ is the best sensitive probe for the inflowing gas with the temperature of $\sim10^4$--10$^5$~K between galaxies and the IGM, and a few cases of inflows have been revealed at high redshifts\cite{Martin2019}. By mapping the Ly$\alpha$ emission in the crucial outskirt region of nearby galaxies, CAFE will directly observe the spatial distribution and kinematics of the inflowing gas, and thereby will test prevailing accretion ideas.

The feedback is normally in the form of  multi-phase outflows, including cold neutral clouds ($10^{3}$ K), warm ionized gas ($10^{4}$~K), and coronal plasma ($10^{5}$ to perhaps $10^{7}$ K). However, only the coronal plasma may transport mass, momentum, energy, and heavy elements into the CGM and IGM. The hot ($>10^{6}$~K) phase can be studied in X-ray emission, and several large X-ray missions have been proposed with enough sensitivities for mapping large scale structures. The best tracers of the gas in the $10^{5}-10^{6}$~K temperature range are the \ovi{} 
doublet, as oxygen is the most abundant metal and the dominant coolant for gas at temperatures of a few times 10$^5$~K\cite{Gnat2007}. Current observations are dominated by absorption studies, but again, this type of observation has very little information about the transverse extent of the absorbing gas clouds. With the \ovi{} emission mapping, the CAFE census would form the foundation for building a physical understanding of galaxy feedback. In addition, the kinematic mapping of the followed deeper CAFE observations and the spectroscopy power of LyRIC will provide insight into the interactions between multi-phase gas in galactic winds and galactic fountains. CAFE will also study cooling, accretion, and feedback in galaxy clusters.

\subsection{Baryon cycle and Energetic Feedback in the Milky Way and Magellanic Clouds}

Due to its closeness, the characterization of the plasma in the Milky Way and the Magellanic Clouds will be the first step to understand the astrophysics of gaseous halos of galaxies. The HVCs revealed by the \hi{} 21cm-line observations and the UV absorption studies are gas flows in Galactic halo\cite{Putman2012}, and thought to be a reservoir to sustain star formation in the Milky Way\cite{Lehner2011}. However, either their origin or fate is still under debate. Their infalling process toward the galactic disk is accompanied by the interaction with the hot gaseous halo\cite{Wakker1999,Tripp2012}, and forms a warm-hot transition layer of 10$^{5-7}$~K that revealed by the absorption features of highly ionized species such as \ovi{}, \nv{}, and \civ{} in the FUV band\cite{Sembach2003}.  Therefore, a census of the line absorption and emission properties will provide valuable information, especially in the Lyman-UV regime that cannot be probed with current facilities. The \ovi{} $\lambda$1032 emission feature in the Milky Way has been detected by FUSE and exhibits intensities of 1900--8600 PU\cite{Otte2006}, which places a sensitivity requirement to the proposed LUV missions. 


Besides that, hot bubbles including supernova remnants and superbubbles surrounding massive star formation regions are the sites of the stellar feedback process. The ionizing photons, mechanical energy, and synthesized metals shed or ejected from newly formed stars shape ambient medium, closing the baryon cycle in normal galaxies, or even triggering galactic outflow that influences CGM/IGM in starburst cases\cite{Veilleux2005}. However, a common query on its efficiency is that radiation carries away most of the feedback energy presumably\cite{Katz1992}. The characterization of transition layer between the hot interior and cold gas shell is of unique importance to probe the efficiency issue, because most of the processes that govern the bubble evolution and radiation property, including thermal conduction, radiation momentum transfer, shell fragmentation, hot gas leakage, etc., all happen at the transition layer. Therefore, the radiation property of high ions in the hot bubbles probed with LUV missions will provide the most important piece to the multi-wavelength study of stellar feedback process.

\section{Proposed Missions in China}

\subsection{LyRIC}\label{sec:lyric}

LyRIC is designed to take advantage of the mature technology, long-slit spectrograph (LSS) operating on Chinese Space Station to fulfill the high quality spectroscopic radiation measurements from CGM of nearby galaxies with big angular sizes, HVCs in our Galaxy and M31, and the diffuse sources in our Galaxy and SMC/LMC for the first time. Besides the scientific significance, such a scientific payload will also help us to verify our LUV technologies in space. 

\begin{figure}
    \centering
    \includegraphics{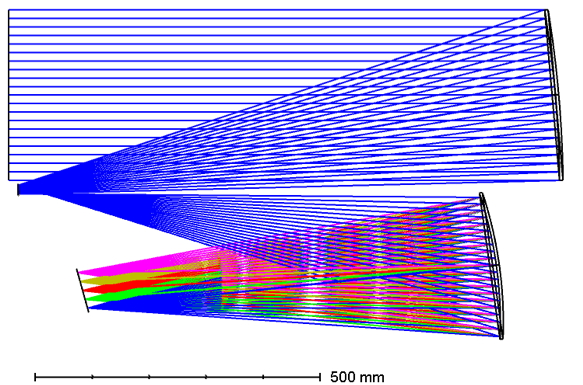}
    \caption{Optical design for LyRIC}
    \label{fig:lyric_opt}
\end{figure}

Limited by the size and weight of the independent external payload operating on CSS, the major specifications for LyRIC is summarized in Table~\ref{tab:lyric_spec}. LyRIC  will have a mirror with aperture less than 30 cm, spectral resolution R$\sim$ 3000, and spatial resolution about 30 arcsec, working in the wavelength range of 91-115 nm for a lifetime longer than 5~years. 

National Astronomical Observatory in China (NAOC) has succeeded in building the prototype LSS and vacuum test facilities. The prototype FUV LSS is sensitive down to 102 nm, which gives us a precious experience. 

\begin{figure}
    \centering
    \includegraphics[width=0.38\textwidth]{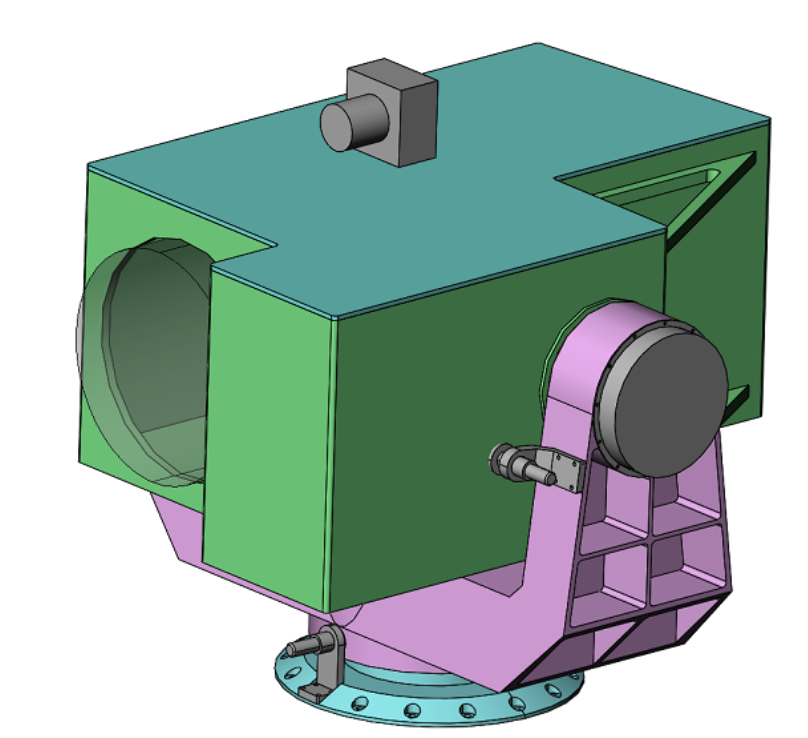}
    \includegraphics[width=0.52\textwidth]{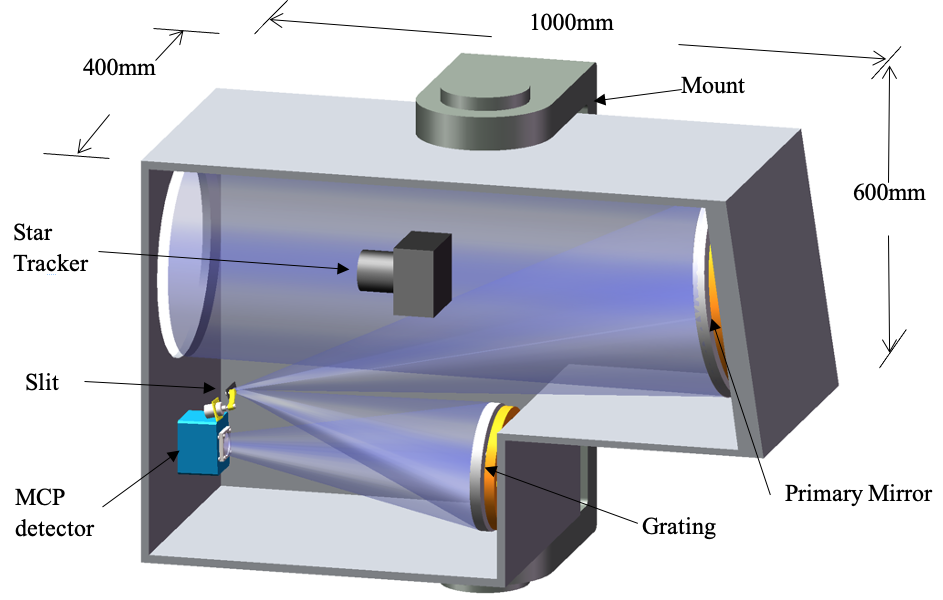}
    \caption{The outlook (left) and inside view (right) of LyRIC instrument.}
    \label{fig:lyric_scr}
\end{figure}

LyRIC adopts a traditional optical design for long-slit spectrograph as shown in Fig.~\ref{fig:lyric_opt}. The primary mirror is an off-axial paraboloidal mirror with a focal ratio of F/3.3. Light collected by the primary mirror is first folded by a small planar mirror and then passes an entrance slit of $18\times 0.2$ mm. A spherical holographic grating with a diameter of 250 mm and focal ratio of F/1.5 is placed behind the entrance slit and refocus the diffracted lights onto a microchannel plate (MCP) detector. The holographic grating has an effective groove density of 3600 g/mm and its first-order diffraction is used. LyRIC has a dimensional FOV of $1^{\circ}\times 0.5^{\prime}$ with $30^{\prime\prime}$ spatial resolution in the cross-diffraction direction. The one-dimensional imaging spectrum covers a wavelength of 91--115~nm and occupies a physical area of $64\times 16$ mm on the MCP detector. The preliminary design for the LyRIC instrument is shown in Fig.~\ref{fig:lyric_scr}.

\begin{table}[htp]
    \centering
    \caption{Major specifications for LyRIC}
    \begin{tabular}{l|l}
      \hline
      \hline
      Aperture   & 300~mm \\
      Wavelength & 91 -- 115~nm \\
      Spectral Resolution & $R\approx3000$ \\
      Instantaneous FOV   & 1$^\circ\times0.5^\prime$ \\
      Angular Resolution  & 30$^{\prime\prime}$ \\
      Payload Envelope     & $1.0\times0.6\times0.4$~m \\
      \hline
      \hline
    \end{tabular}
    \label{tab:lyric_spec}
\end{table}

\subsection{CAFE}\label{sec:cafe}

\begin{figure}
    \centering
    \includegraphics[width=0.6\textwidth]{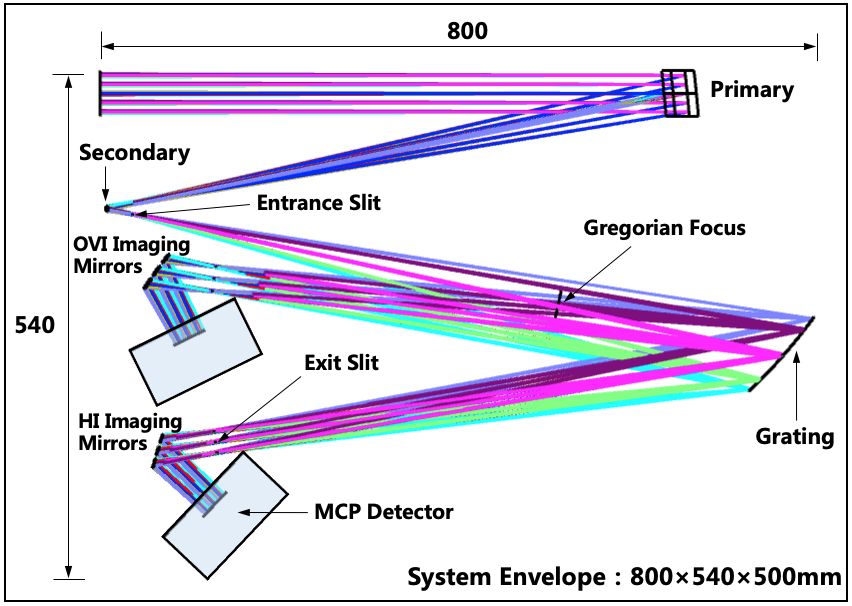}
    \caption{CAFE optical layout}
    \label{fig:cafe_opt}
\end{figure}

\begin{figure}
    \centering
    \includegraphics[width=0.37\textwidth]{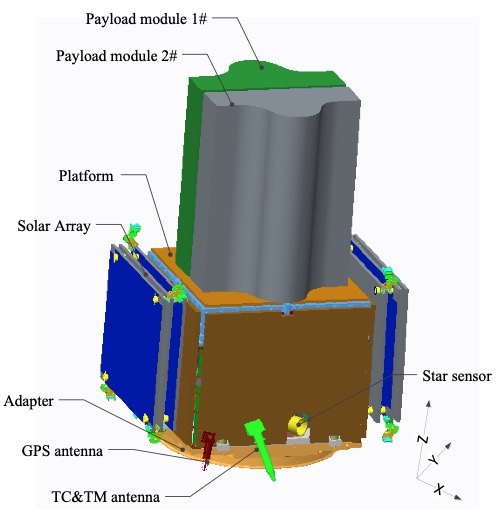}
    \includegraphics[width=0.61\textwidth]{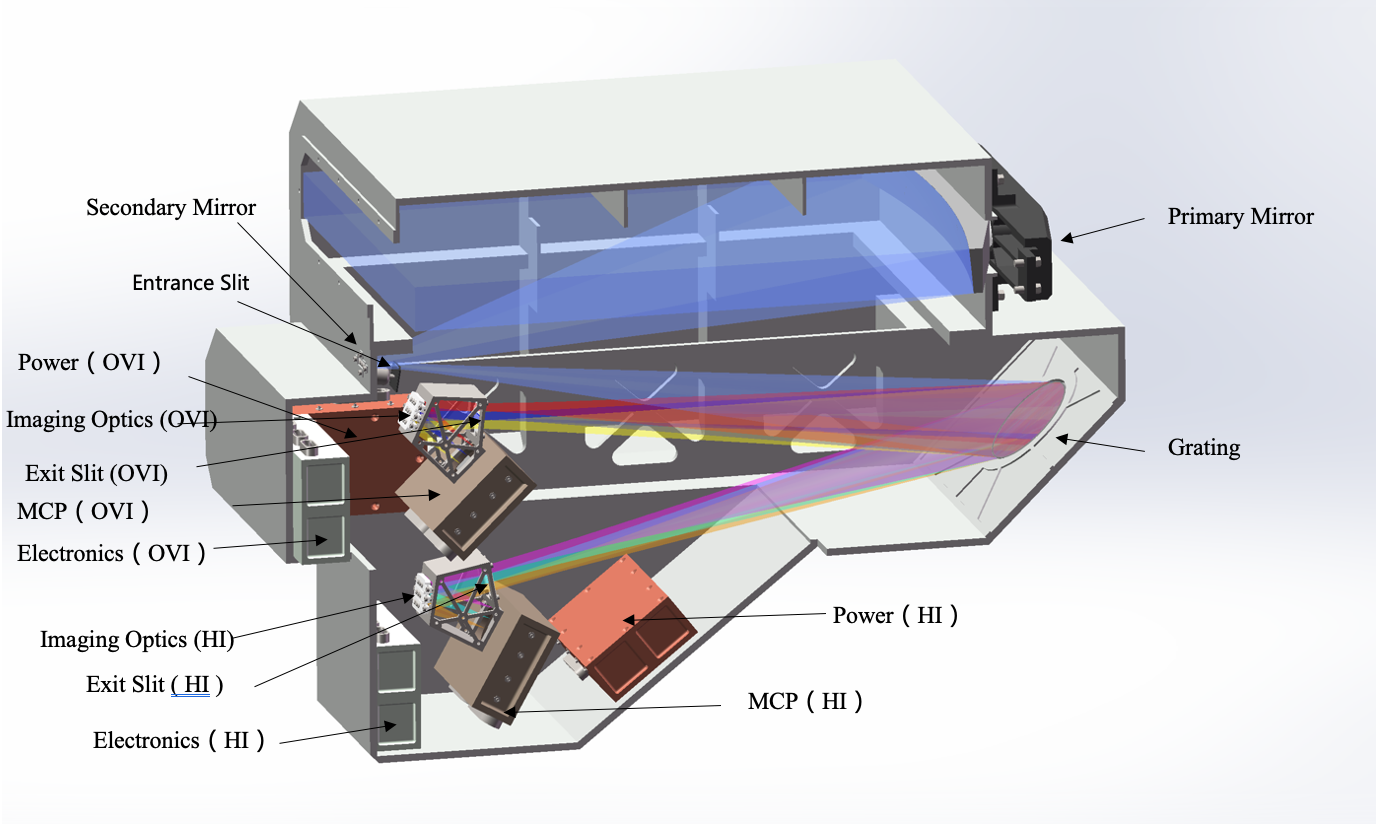}
    \caption{The outlook (left)  and inside view (right)  of CAFE instrument.}
    \label{fig:cafe_apr}
\end{figure}

CAFE will use a novel optical design\cite{Lou2019} and an advanced photon-counting MCP detector (see section \ref{sec.mcp}) for two identical multi-channel narrow band imagers, mainly observing the \ovi{} and \hi{} emission lines at the redshifts of the pre-selected target galaxies in the local Universe (z=0-0.05) with FOV about $20^\prime \times 20^\prime$ and spatial resolution at least $30^{\prime\prime}$. Our 3-yr baseline mission has low inclination and low earth orbit, mounted on an agile spacecraft, with a survey mode ($R = \lambda/\Delta\lambda = 500$) and a mapping mode (R=2000). CAFE must achieve a $3 \sigma$ detection limit of 500 PU and the instrument must allow co-adding to achieve higher sensitivity. The major specifications are summarized in Table~\ref{tab:cafe_spec}. The outlook and inside view of CAFE instrument are shown in Fig~\ref{fig:cafe_apr}.

The optical layout is shown in Fig~\ref{fig:lyric_opt}. An offset Gregorian telescope collects the incident light and feeds its converging beams into a monochromator system. The monochromator consists of an entrance slit, a highly dispersive grating, and six exit slits defining six narrow-band imaging channels. The entrance slit is located on the image of the primary formed by the Gregorian system immediately after the secondary. Therefore, the entrance slit defines the usable aperture of the primary. Note that such a design is significantly different from a traditional long-slit spectrograph design where the entrance slit is located on the focal plane. The grating is placed at some distance after the Gregorian focus. It disperses the incident light according to the wavelength and at the same time forms another pupil plane along its exit directions. Six exit slits are placed on this pupil plane to define six distinct wavelength channels. The three \ovi{} channels are centered at 103, 104.8 and 106.6 nm, while the three \hi{} $\rm Ly\alpha$ channels are centered at 125.0, 126.2 and 127.4 nm, respectively. The central wavelengths of the imaging channels are tunable by rotating the exit slits from z=0 to 0.05 (\ovi{}) while from z=0.03 to 0.05 (\hi{} $\rm Ly\alpha$ for the low orbit design), imaging mirrors and MCP detector together along a circular path centered at the grating. Pupils of different channels are well separated in the dispersive direction so that the light from any field of one channels does not overlap with any field of another channel. The bandwidth (or spectral resolution) of each channel is adjustable by varying the width of the entrance and exit slits. Finally, each exit slit is followed by an imaging mirror which forms the full-field ($20^\prime \times 20^\prime$) image of a particular channel on a planar MCP detector. The three \ovi{} channels share the same detector while the three \hi{} Ly$\alpha$ channels share another detector.


\begin{table}[htp]
    \centering
    \caption{Major design specifications and constraints for the CAFE imager}
    \begin{tabular}{l|l}
      \hline
      \hline
      {\bf Parameter}            & {\bf Specification} \\
      \hline
      Wavelength Coverage        & 102 -- 130~nm \\
      Number of Imaging Channels & 3 \ovi{} + 3 \hi{} Ly$\alpha$ \\
      Spectral Resolution        & 500/1000/2000 \\
      Field of View              & 20$^\prime \times 20^\prime $ \\
      Spatial Resolution         & $<30^{\prime\prime}$ \\
      Sensitivity                & 500~PU (3$\sigma$, 10$^4$~s, $R=500$) \\
      Payload Dimension Constraint & $<1000\times500\times500$~mm \\
      MCP Size and Resolution    & $39\times39$~mm, 20~$\mu$m \\
      \hline
      \hline
    \end{tabular}
    \label{tab:cafe_spec}
\end{table}

\begin{figure}
    \centering
    \includegraphics{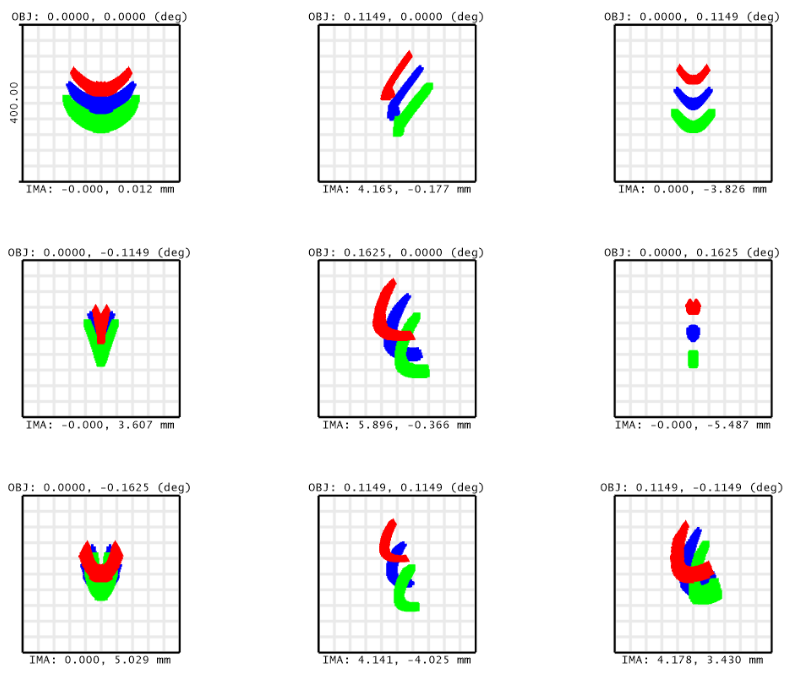}
    \caption{CAFE spot diagram for the 104.8 nm imaging channel}
    \label{fig:cafe_spot}
\end{figure}

The grating in the CAFE optics has to provide sufficient chromatic dispersion to separate images of different wavelength channels. It also creates a pupil plane after the grating to place the exit slits as well as helps to control the optical aberrations. Holographic gratings are favored because they allow the use of a spherical substrate. In our design, the grating is a third-order holographic concave grating with a spherical substrate and an effective density of 3350~g/mm. The holographic recording geometry is optimized to achieve the desired imaging and spectral performances, while the groove profile is optimized to maximize grating efficiency for the third-order diffraction. The typical spot diagram of the imaging channels is shown in Fig.~\ref{fig:cafe_spot}. The 80\% energy encircle diameter is less than 21 arcsec for all channels and across the full fields of $20^\prime \times 20^\prime$, which meet the specification of 30 arcsec by 30\% margin.

\section{Enabling Technologies}
\subsection{Mirror Coatings}\label{sec.coating}
It has been found that Al coating has a high reflectance in the far ultraviolet range and good adhesion to the substrate, which is the best choice for the preparation of high-quality mirrors. However, the surface of the Al film is easily oxidized to form a thin layer of $\rm Al_{2}O_{3}$, which has a large absorption below 160 nm. Much work have been done to improve the reflectivity by plating a protective layer\cite{Hass1956,Madden1963}. A fluoride is well suited as protective coating for the Al film because of its low intrinsic absorption, like lithium fluoride (LiF), magnesium fluoride ($\rm MgF_{2}$) and $\rm AlF_{3}$\cite{Hunter1971,Angel1961,Keski-Kuha1999}. Al coating protected with MgF$_{2}$ has been used in HST, and LiF protected Al coatings have been used in FUSE, with the reflectivity about 50\% at 98-118.7 nm\cite{Bianchi2012,Moos2000}. The capabilities of sensitive instruments were limited by the poor performance of conventional mirrors. The NASA TFCL has produced Al+eLiF and Al+eMgF$_{2}$ mirror coatings by heating the substrate to $\sim$220 $^{\circ}$C during the deposition process, which increase the reflectivity up to 70\% at 110 nm (Al+eLiF), 80\% at 115 nm ($\rm Al\times eMgF_{2}$), closer to the theoretical maximum\cite{Quijada2012,Quijada2014}. 

\begin{figure}
    \centering
    \includegraphics[width=0.5\textwidth]{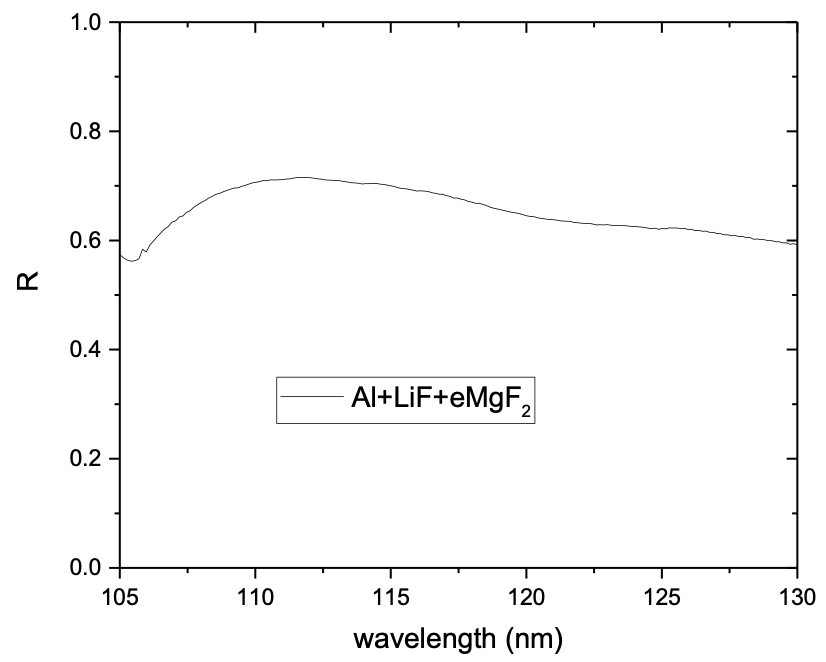}
    \caption{The reflectivity of Al+LiF+eMgF$_{2}$}
    \label{fig:cafe_coating}
\end{figure}

Our laboratory has done a lot of experiments to study Al mirror coatings, prepared Al+eMgF$_{2}$ at different temperatures, and characterized the variety of surface and reflectivity after 30 months. Al+eMgF$_{2}$ coating showed higher reflectivity above 110 nm prepared at 200 $^{\circ}$C and 300 $^{\circ}$C, with lower roughness and better stability. Because LiF is the first choice of overcoat at the short wavelength end of the spectrum, whose absorption edge is at 102 nm.  We prepared Al+eLiF at 300 $^{\circ}$C, it was found that the reflectivity dropped about 40\% in 2 months that stored in 40\% relative humidity (RH), which is more quicker than that stored in the 20\% RH. Therefore, it is necessary to add a protective overcoat layer to reduce the hygroscopic sensitivity of the Al+eLiF coating, The $\rm Al+LiF+eMgF_{2}$ mirror coating was prepared by heating the substrate to $\sim{}200$~$^\circ$C during the deposition process, and the measured reflectivity is shown in Fig.~\ref{fig:cafe_coating}. 

It is reported that the reflectivity can be effectively improved when the Al+eLiF mirror coating overcoated by a thin AlF$_{3}$ film deposited by atomic layer deposition (ALD)\cite{Fleming2017}. And the SiC seems to be a protective layer material worth considering, which shows a reflectivity about 40\% at LUV. We will focus on the study of the protective overcoat layer on Al+eLiF coatings in the future, like AlF$_{3}$, SiC and so on, as well as the performance in different environments.

\subsection{Grating}\label{sec.grating}
As a critical component, varied line-space concave gratings (VLSCG) have been widely used in soft X-ray to visible spectrum instruments, including FUSE mission. FUSE operates in a band of 90-120 nm, which contains two channels, each with two VLSCGs, with working in bands 91 -- 103 nm{} and 104 -- 120 nm, respectively.


CAFE is composed of an incident slit, a concave grating and six exit slits (see Session \ref{sec:cafe}). With the VLSCG, beams with the same wavelength from diﬀerent ﬁeld of view converge on the same exit slit, and the exit slits for different channels are distributed along a ring centered on the grating. Similar to the incident slit, the exit slit is placed at the exit pupil surface formed by the concave grating, and the width of the slit is used to limit the spectral bandwidth. Different wavelengths are separated by the concave grating with suﬃciently high dispersion capability. On the other hand, the grating is also used as an imaging element to image the incident slit and form an exit pupil surface for diﬀerent central wavelengths, while reducing the aberration of the system. In addition, limited space puts a constraint on the size of the system envelope, thus imaging optical elements including gratings have relatively short focal length and small F number.

Experiences in the design and fabrication of VLSCGs in soft X-ray band (0.5--6 nm{})\cite{Chen2015} and large scale multilayer dielectric grating via interference lithography and ion-beam etching techniques\cite{Qiu2018} enable a VLSG with central density of 3350 gr/mm, size of 112 mm (L)$\times$118 mm (H) . The design of the concave grating for CAFE has been completed so far. The influence of alignment errors in holographic optics path on groove density distribution will be systemically analyzed. The grating patterns are recorded into photoresist using aspheric wave-front recording optics. The groove depth and duty cycle of photoresist grating are adjusted through RIE with O$_{2}$ for high diffraction efficiency. At last, the grating will be coated with a layer of Al+LiF+eMgF$_{2}$ for high reflectivity. In addition, we plan to transfer the photoresist patterns into fused silica substrate by reactive ion beam etching for a higher diffraction efficiency, which is predicted up to 40\%.

The design of the holographic recording configuration and fabrication process of the VLSCG for CAFE are in process, with a targeted delivery date in the winter of 2021.

\subsection{Advanced MCP Detectors}\label{sec.mcp}

At the “Institut für Astronomie und Astrophysik T\"ubingen” (IAAT), advanced imaging and photon counting MCP detectors\cite{Conti2018} are developed for CAFE. The detectors make use of gallium nitride (GaN) as photocathode material and comprise a stack of two MCPs and a coplanar cross-strip (CS) anode with advanced readout electronics.

The GaN photocathode is adopted to fulfill the requirement of high quantum efficiency for the CAFE detector. For the 124 -- 130 nm band the detector will be sealed with a MgF$_2$ window on which a Cs-activated photoactive layer of GaN is grown. For the 102-108 nm line emission the transmission rate through a window would be too low and a door will be used instead of the window. In this “open design” we grow GaN directly on the MCP. The sealing of the detector with a window or a door is to prevent a degradation of the photocathode. The effect of temperature on the growth of GaN on MgF$_2$ by plasma-assisted molecular beam epitaxy (PA-MBE) has been investigated\cite{Meyer2020a,Meyer2020b}.  We have also optimized the thickness of the GaN layer and studied the effect of Mg-doping concentration. 

Long-life MCPs (by PHOTONIS Technologies) or ALD MCPs will be used to achieve the goal of the desired low dark rate. The coplanar CS anode with 64 strips in each direction was designed by IAAT in cooperation with the manufacturer (VIA electronic GmbH, Hermsdorf, Germany).  Each of the 128 anode strips is linked to an input channel of a special chip, an ASIC developed by the Max Planck Institute for Nuclear Physics in Heidelberg for the LHCb experiment\cite{Agari2004}.  The chip is a charge preamplifier and shaper. It provides sampling frequencies of up to 40 MHz, a very small equivalent noise charge of only a few hundred electrons and can store up to 160 samples for each channel\cite{Lochner2006}. We have already defined promising centroiding algorithms for the CS-anode signals.

\section{Status}
LyRIC has been proposed to the China Manned Space Engineering Office in 2019. We are highly recommended among the astronomical instruments. Since then, we have made a couple of modifications according to the requirements of CSS accommodations and interfaces. Further optimization is anticipated to compromise among science, technology, as well as the requirement of the external payload operating on CSS. 

CAFE is at the stage of Key Technologies Investigation in China (2016-2021). The optical elements are ready to be fabricated by Nanjing Institute of Astronomical Optics and Technology of CAS. Mirror coatings will be completed in Tongji University in Shanghai. Grating will be made in the Great grating group in University of Science and Technology of China. The mechanical structures will be made by Xi'an Institute of Optics and Precision Mechanics of CAS. MCP detectors will be developed by Institute for Astronomy and Astrophysics, Universit\"at T\"ubingen in Germany. Purple Mountain Observatory will build the Lyman UV lab at Xianlin Campus to do the assembling and verification. After the key technology development period, a positive outcome will allow us to be ready to the competition of Phase A studies for Space Missions in China. The final design of CAFE will be subject to mission opportunities in China in the future.  

In addition to undertake instrument performance analysis and to produce simulated surveys for mission- and science studies, we are carrying out the end-to-end simulation with the Monte Carlo based simulator SIXTE\cite{Wilms2014,Dauser2019}, in collaboration with the software developer at the Remeis Observatory of der FAU Erlangen-N\"urnberg in Germany.

\acknowledgments 
The scientific program described here was originally developed by members of CAFE and LyRIC teams including: Q.S. Gu, P. Zhou, Y. Chen, Z. Li, J. Zhou, X. Hou, X. Luo (NJU), Y. Gao, X.Z. Zheng, X. Kang, L. Feng, X.J. Jiang, C. Ge, J. Ji, S. Wang, G. Chen,
S. Jin, X. Zhou, Y. Zhang, X. Long, W. Shan (PMO), F. Cheng, X. Kong, G. Liu (USTC), J. Wang, F. Zhang
(Xiamen U.), L. Ho (KIAA), X. Fang, L. Gao, J. Liu (NAOC), Z. Cai (Tsinghua U.),
H.G. Xu, J. Shen (Shanghai Jiaotong U.), L. Hao (SHAO), J.R. Mao (YNOC) from China, and our international collaborators: M. Baes (U. Gent, Belgium), M. A. Barstow (U. Leicester, UK), R. de Grijs (Macquarie U., Australia), M. Hayes, R. Lallement (GEPI, France), F. Nicastro (Istituto Nazionale di Astrofisica, Italy), P. Richter (University of Potsdam, Germany), D. Valls-Gabaud (GEPI, France), J. Bregman, J. Li, Z. Qu (U. Michigan, USA), Q. D. Wang (UMASS, USA), Kinwah Wu (MSSL, UK), and J. Wilms (FAU Erlangen-N\"urnberg, Germany).

L. Ji acknowledges support from NSFC grant U1531248 and CAS International Cooperation Key Project grant 114332KYSB20180013.

\bibliography{report} 
\bibliographystyle{spiebib} 

\end{document}